\documentclass[10pt]{article}
\def\be{\begin{equation}}
\def\ee{\end{equation}}
\def\bea{\begin{eqnarray}}
\def\eea{\end{eqnarray}}
\def\pd{\partial}

\begin{document}
\title{Comments on Galileons}
\author{David Fairlie$\footnote{e-mail: david.fairlie@durham.ac.uk}$\\
\\
Department of Mathematical Sciences,\\
         Science Laboratories,\\
         University of Durham,\\
         Durham, DH1 3LE, England}
\maketitle
\begin{abstract}
The recent progress in the study of Galileons, i.e. equations of second order with an action invariant under a Galilean transformation 
is related to work on  `Universal Field Equations' \cite{dbfgov} which are second order equations arising by an 
iterative procedure from arbitrary Lagrangians of weight one in their first derivatives. It is pointed out that the Galileon is 
simply a Kaluza-Klein reduction of a Universal Field Equation. An implicit solution to the equation of motion 
is presented, and a class of explicit solutions pointed out. The multi-field extensions of both types of equations are derived
from  a first order formalism, which is simply the substantive derivative of fluid dynamics. 
\end{abstract}
 
\section{Introduction}
Recently there has been a spate of articles on the subject of Galileons, i.e. equations, second order in derivatives of a field $\pi$ with an action
invariant under the infinitesimal transformation
\be \pi(x) \mapsto \pi(x) +c +v_\mu x^\mu+\pi(x)v^\mu\pd_\mu\pi(x)\label{oneone}\ee
with $x^\mu$  space-time co-ordinates, $v_\mu$ a constant  vector and $c$ a constant.
\cite{nicolis} \cite{derahm}\cite{deser}. 
In the literature two versions of the Lagrangian in 3+1 dimensions are proposed; one starting with the Lagrangian
\be {\cal L}_2 = -\sqrt{1+(\partial\pi)^2},
\ee
which is invariant up to a divergence under the transformation (\ref{oneone}) 
and continuing with a heierarchy of  increasingly complicated Lagrangians  \cite{derahm}, 
ending with
\be
{\cal L}_5 =-\gamma^2([\Pi]^3+2[[\Pi^3]-3[\Pi][\Pi]^2)-\frac{3}{2}\gamma^4(4[\Pi][\pi^4]-4[\pi^5]-2([\Pi]^2-[\Pi^2])[\pi^3]).
\ee
See also \cite{trodden1}.The other starts with the Klein Gordon Lagrangian \cite{trodden2}.
Here the notation $\Pi_{\mu\nu} =\partial_{\mu}\partial_{\nu}\pi$; $[\Pi]^n]={\rm trace}(\Pi^n)$ and $[\pi^n] =\partial\pi.\Pi^{n-2}.\partial\pi$;
i.e $[\pi^3] =\partial_\mu\pi\partial^\mu\partial^\nu\pi\partial_\nu\pi.$ The factor $\gamma$ is the inverse of $\sqrt{1+(\partial\pi)^2}$.
The resulting equation of motion is ${\cal E}_5\,=\,0 $ where
\be
{\cal E}_5= \gamma^6\left([\Pi]^4 -6[\Pi]^2[\Pi^2]+8[\Pi][\Pi^3]+3[\Pi^2]^2 -6[\Pi^4]\right).\label{hess4}
\ee
We shall discuss this procedure in the light of previous work on iterated Lagrangians, and relate it to these recent studies.
This equation of motion can be very much more simply expressed thanks to a result on determinants, which is published in Lowell Brown's book,
\cite{lb}  but was independently found by myself; here illustrated for 4x4 matrices. Consider the determinant 
\be
\det\left|\begin{array}{ccccc}
     x^4&x^3&x^2&x&1\\
     T_4&T_3&T_2&T_1&4\\          
     T_3&T_2&T_1&3&0\\ 
     T_2&T_1&2&0&0\\
     T_1&1&0&0&0
\end{array}\right|\label{det}
\ee 
where $T_n ={\rm trace}(T^n)$, and $T$ is a matrix of dimension 4x4.
The expansion of this determinant is just $4!$ times the characteristic polynomial of the matrix $T$, and the coefficient independent of $x$
is just the determinant of $T$.
\[ \det(T) =-6*T_4+8*T_3*T_1+3*T_2^2-6*T_2*T_1^2+T_1^4\]
But, if $T_{\mu\nu} =\frac{\partial^2}{\pd_{\mu}\partial_{\nu}}\pi=\Pi_{\mu\nu}$, this is just the operative factor in ${\cal E}_5$, so the equation of motion 
simply obtained by setting the Hessian $\displaystyle {\det|\Pi_{\mu\nu}|}=0.$ This is also called the homogeneous
 Monge-Ampe\`re equation and a general  solution has been given by A.N. Leznov and myself \cite{leznov}. I recently recognised an equivalent, 
and perhaps more elegant solution in a book of T.H. Chaundy, quite some time ago \cite{chaundy}.
\subsection{Chaundy's solution}

The principle is illustrated by the 3x3 case. To solve
\begin{equation}
\det\left|\begin{array}{ccc}
          \phi_{xx} & \phi_{xy}&\phi_{xz}\\                    
        \phi_{xy}&\phi_{yy}&\phi_{yz}\\
        \phi_{xz}&\phi_{yz}&\phi_{zz}
\end{array}      \right|\,=\,0,\label{MA}
\end{equation}
choose four arbitrary functions $f(u,v),\ g(u,v),\ h(u,v),\ k(u,v)$ constrained by the three relations
\bea
&& xf(u,v)+yg(u,v) +zh(u,v)+k(u,v)=\phi(x,y,z)\\ 
&& x{ f(u,v)_u}+y{ g_u(u,v)} +z h_u(u,v)+ k_u(u,v) = 0\\
&& x{ f_v(u,v)}+y{ g_v(u,v)} +zh_v(u,v)+k_v(u,v) = 0.
\eea
Here subscripts denote partial differentiation with respect to $u,\ v$.
Then the implicit solution of these equations for $\phi(x,y,z)$ is a solution to (\ref{MA}).
This is easy to see; the three equations imply
\be 
f(u,v) = \frac{\pd\phi(x,y,z)}{\pd x};\ \ g(u,v) = \frac{\pd\phi(x,y,z)}{\pd y};\ \ h(u,v) = \frac{\pd\phi(x,y,z)}{\pd z}.
\ee
Since the right hand sides are three functions of two variables, there must be a relationship amongst them; i.e.
there is a functional relationship among all the first derivatives of $\phi$, say $F(\phi_x,\phi_y,\phi_z)$.
As we know this means that $\phi$ satisfies the Monge-Amp\`ere equation. 
This is easily seen from the assumption of a relationship of the form
\[ F(\phi_x,\phi_y,\phi_z)=0\]
Differentiation with respect to the variables $x,y,z$ respectively gives the three relations
\bea
\frac{\pd F}{\pd\phi_x}\phi_{xx}+\frac{\pd F}{\pd\phi_y}\phi_{yx}+\frac{\pd F}{\pd\phi_z}\phi_{zx}&=&0\\
\frac{\pd F}{\pd\phi_x}\phi_{xy}+\frac{\pd F}{\pd\phi_y}\phi_{yy}+\frac{\pd F}{\pd\phi_z}\phi_{zy}&=&0\\
\frac{\pd F}{\pd\phi_x}\phi_{xz}+\frac{\pd F}{\pd\phi_y}\phi_{yz}+\frac{\pd F}{\pd\phi_x}\phi_{zz}&=&0.
\eea
The eliminant of these equations is simply the Monge-Ampe\`re equation (\ref{MA}).
We can also verify directly this by constructing the determinant of the matrix of second derivatives
\begin{equation}
\det\left|\begin{array}{ccc}
          f_uu_x+f_vv_x&f_uu_y+f_vv_y&f_uu_z+f_vv_z\\                    
          g_uu_x+g_vv_x&g_uu_y+g_vv_y&g_uu_z+g_vv_z\\  
          h_uu_x+h_vv_x&h_uu_y+h_vv_y&h_uu_z+h_vv_z  
        \end{array}      \right|\label{MB}
\end{equation}
Evaluation of this determinant gives zero. It is also easy to see, by a similar technique that any homogeneous
function of weight one, i.e which satisfies
\[ x\frac{\pd\phi}{\pd x}+ y\frac{\pd\phi}{\pd y}+ z\frac{\pd\phi}{\pd z}= \phi\]
will also satisfy the equation (\ref{MA}) by a similar technique of differentiation by $x,y,z$ in turn.
Interestingly, it is easy to see in this case that the Galilean transformation (\ref{oneone}),
not now necessarily infinitesimal also satisfies the equation (\ref{MB}) as, apart from the irrelevant constant, the additional terms 
are also of weight one, if $\pi$ is.
One might worry as to whether the equality of the two different ways of evaluating mixed second derivatives 
introduces further constraints; for example
\bea
 \frac{\pd^2\phi}{\pd x,\pd y}& = &\frac{\pd f(u,v)}{\pd u}\frac{\pd u}{\pd y}+\frac{\pd f(u,v)}{\pd v}\frac{\pd v}{\pd y}\\
\frac{\pd^2\phi}{\pd y,\pd x}& = &\frac{\pd g(u,v)}{\pd u}\frac{\pd u}{\pd x}+\frac{\pd g(u,v)}{\pd v}\frac{\pd v}{\pd x}.
\eea
It turns out that the equality of the two expressions is automatically satisfied, and goes back to the equality of the mixed derivatives
of $f,g,\dots.$ etc. Putting it more simply $f_u u_y+f_v v_y\,=\, g_uu_x+g_vv_x$.
This is then a solution to the 3-dimensional Galileon equation
 \be \det\left|\frac{\pd^2\phi}{\pd_\mu\pd_\nu}\right| = 0,\label{laplace}
\ee
since this is just 6 times the determinant (\ref{MB}) in the case where $\mu,\ \nu$ run from 1$\dots$3. If the range is 1$\dots n$, 
the determinant decomposes into the sum of ${\displaystyle\frac{n!}{(n-3)!}}$ Hessian determinants, and the same form of solution
 will hold by the elimination of $u,\ v$ from the equations
 \bea
\phi(x_\mu)&=& \sum_{\mu=1\dots n+1} x_\mu f^\mu(u,v)\\
&&  \sum_{\mu=1\dots n+1} x_\mu \pd_u f^\mu(u,v)\,=\, 0\\
&&  \sum_{\mu=1\dots n+1} x_\mu \pd_v f^\mu(u,v).\,=\,0.
\eea
 The  construction in general  can be easily inferred, and the analogous treatment of the solution of a related 
equation,  described in the following section has been already described \cite{fai3}.  

\section{Iteration of Lagrangians}
This version of a Galileon equation is actually simply a Kaluza Klein reduction of what J.Govaerts, A. Morozov and I called
a `Universal Field Equation'  \cite{dbfgov}.  We called it this because, starting with any Lagrangian dependent only upon $\pi$ and
 its first derivatives,and homogeneous of weight one in these derivatives, i.e.
\[ \sum\frac{\pd {\cal L}}{\pd(\pd_\mu\pi)}{\pd_\mu\pi}={\cal L},\]
subject to an iterative procedure, gives a UFE which is independent of the form of the initial Lagrangian. 
  This obtained by iterating the Euler operator ${\cal E}$ acting on the Lagrangian
\be
{\cal E}=-\frac{\pd}{\pd\pi}
 +\pd_i \frac{\pd}{\pd\pi_i}-\pd_i\pd_j\frac{\pd}{\pd\pi_{ij}}\dots\label{elop}
\ee
(In principle the expansion continues indefinitely  but it is sufficient for
our purposes to terminate at the stage of second derivatives  $\phi_{ij}$,
since it turns out that the iterations do not introduce any derivatives
higher than the second).
We may start with a Lagrangian say, $\displaystyle{{\cal L}=\sqrt{\sum_{\mu=1\dots 5}(\pd_{\mu}\pi)^2}}$.
Then the 5 fold iteration
\be
 {\cal E}{ \cal L} {\cal E}{\cal L}{\cal E}{\cal L },\cdots,{\cal E}{\cal L}\label{iter}
\ee
where each Euler operator acts on everything to the right yields the Universal
Field Equation
\be
\det\left|\begin{array}{cccccc}
          0&\pi_{x_1}&\pi_{x_2}&\pi_{x_3}&\pi_{x_4}&\pi_{x_5} \\
          \pi_{x_1}&\pi_{x_1x_1}&\pi_{x_1x_2}&\pi_{x_1x_3}&\pi_{x_1x_4}&\pi_{x_1x_5}\\
           \pi_{x_2}&\pi_{x_2x_1}&\pi_{x_2x_2}&\pi_{x_2x_3}&\pi_{x_2x_4}&\pi_{x_2x_5 }\\
           \pi_{x_3}&\pi_{x_3x_1}&\pi_{x_3x_2}&\pi_{x_3x_3}&\pi_{x_3x_4}&\pi_{x_3x_5}\\
          \pi_{x_4}&\pi_{x_4x_1}&\pi_{x_4x_2}&\pi_{x_4x_3}&\pi_{x_4x_4}&\pi_{x_4x_5}\\
           \pi_{x_5}&\pi_{x_5x_1}&\pi_{x_5x_2}&\pi_{x_5x_3}&\pi_{x_5x_4}&\pi_{x_5x_5}\end{array}\right|\,=\,0.
\label{univ}
\ee
This equation possesses some interesting invariance properties; First of all it is signature blind; being the same in a Euclidean
space-as a Lorentzian space-time. Also, any function of a solution is also a solution! This may be verified directly, but it  arises
from the solution procedure explained in \cite{fai3}.
Now in a  Kaluza Klein reduction we may choose a gauge in which $\displaystyle{\frac{\pd\pi}{\pd x_5}\,=\,1}$ . In this case the last row and column of
(\ref{univ})  have only two nonzero elements in the (1,6) and (6,1) positions, and these are unity. The resulting determinant 
simply reproduces (\ref{MA}), the Hessian.
\section{Connection with earlier results}
These results are easily understood as examples of the main results of \cite{dbfgov} quoted here;
\vspace{5pt}
{\small `In two recent papers \cite{dbfgov,gov3}, the first with A. Morozov,
hierarchies of Lagrangian field theories with the following
properties were introduced.
\begin{enumerate}
\item{i)} In any of these hierarchies, the Lagrangian at any given level
-- except of course at the first level -- is essentially proportional to the
equations of motion of the Lagrangian at the previous level (hence the name
Euler hierarchies).
\item{ii)} The proportionality factor mentioned in i) is essentially the very
first Lagrangian in the hierarchy.
\item{iii)} In any of these hierarchies, Lagrangians depend on fields only
through their first and second derivatives, but {\it not\/} on
derivatives of {\it higher\/} order nor on the fields themselves. The first
Lagrangian only depends on first derivatives of the fields. The dependence of
each of the other Lagrangians on second derivatives is
multilinear, and of order equal to the number of times an equation
of motion has been taken to reach that level in the hierarchy.
\item{iv)} All these hierarchies are {\it finite\/}, {\it i.e.} the
iterative procedure implied by i) -- iii) terminates after a {\it finite\/}
number of steps.
\item{v)} For each hierarchy, the last non trivial equations of motion
are universal, namely, up to a
factor, they are {\it independent\/} of the initial Lagrangian out of
which the hierarchy is constructed. The associated infinite number of
conserved charges -- corresponding to the freedom in the choice of
initial Lagrangian -- suggests the possible integrability of these
universal equations (equations of motion are indeed always current
conservation equations for Lagrangians without an explicit dependence
on fields).
\end{enumerate}

\vskip 10pt
Specifically, hierarchies with these properties were shown
to exist in the following cases:
\begin{itemize}
\item{} a single field $\phi$ in $d$ dimensions, with an arbitrary
initial Lagrangian (function of first derivatives only)\cite{dbfgov},
\item{} a single field $\phi$ in $(d+1)$ dimensions, the initial
Lagrangian now being an arbitrary homogeneous weight one function of its
arguments \cite{dbfgov}
\item{} $(d+1)$ fields $\phi^a$ in $d$ dimensions, with an
arbitrary reparametrisation invariant initial Lagrangian \cite{gov3}.
\end{itemize}
The hierarchies associated with these three cases terminate after $d$
steps, with the following universal equations:
\begin{enumerate}

\item{}
\be\det{\phi_{ij}} = 0,\label{one}
\ee
\item{}
\be \det\pmatrix{0&\phi_j\cr\phi_i&\phi_{ij}\cr} = 0,\label{two}\ee
\item{}
\be\det({J_a\phi^a_{ij}}) = 0.\label{three}\ee

\end{enumerate}
Here, $\phi_i$ and $\phi_{ij}$ denote the partial derivatives
$({\pd\phi}/{\pd x_i}),\ ({\pd^2\phi}/{(\pd x_i\pd x_j)})$ of the field
$\phi$ with respect to the $d$ or $(d+1)$ coordinates $x_i$. (The same
applies of course to the fields $\phi^a$, and obviously the indices $i$ and
$j$ in the equations above refer to lines and columns respectively of the
corresponding matrices. The usual summation convention over repeated indices
is assumed throughout). In (\ref{three}), the quantities $J_a$ are the Jacobians
\[J_a =
(-1)\sp d\epsilon_{ab_1b_2\cdots b_d}\phi\sp {b_1}_1\phi\sp {b_2}_2\dots
\phi\sp {b_d}_d.'\]
}

\vspace{5pt}
 Equation (\ref{two}) is a generalisation of the original two dimensional Bateman
equation \cite{dbfgov} (corresponding to (\ref{two}) with $(i=1,2)$) which is also
known to be integrable \cite {dbfgov}. Finally, equation (\ref{three}) is a generalisation to a
$(d-1)$-dimensional membrane in a $(d+1)$-dimensional spacetime of the
(universal) equation of motion for a parametrised particle in a flat two
dimensional spacetime (corresponding to $d=1$), the latter clearly being
also integrable. Note that (\ref{three}) includes a universal equation for a string
theory in three dimensions ($d=2$), and a universal equation for a membrane
theory in four dimensions ($d=3$).

Remarkably, the three classes of universal equations above are invariant under
arbitrary linear $GL(n)$ transformations in the variables $x_i$ as
well as in the fields $\phi^a$, even though neither the initial nor the
successive Lagrangians in the corresponding hierarchies would generally
possess these symmetries. The equations above thus provide examples of
equations of motion admitting an {\it infinite} number of Lagrangians,
with symmetry properties that these Lagrangians need not possess.

\subsection{Other starting points}
In a second paper Hinterbickler et al \cite{trodden2},
the authors start from a simple massless Lagrangian, ${\cal L}_2=[\pi^2]$ in 4 dimensional space-time and generate a sequence of
Lagrangians with second order equations of mmotion. These actually follow from the iterative procedure described above;
and are obtained by setting each ${\cal E}_j=0,$ where
\bea
{\cal E}_2 &=& -2[\Pi]\\
{\cal E}_3 &=& -3([\Pi]^2-[\Pi^2])\\
{\cal E}_4 &=& -2([\Pi]^3-[\Pi^3]-3[\Pi][\Pi]^2)\\
{\cal E}_5 &=& -\frac{5}{6}([\Pi]^4-6[\Pi^4]-8[\Pi][\Pi]^3+3[\Pi^2]^2).
\eea
The last of these, ${\cal E}_5=0$, is identical, up to an irrelevant factor to equation (\ref{hess4}) !.The other equations are simply
given (again up to an irrelevant numerical factor) by
\be {\cal E}_{n+1} = \det_n\left|\frac{\pd^2\pi}{\pd x_\mu\pd x_\nu}\right| \label{hessn}
\ee
where the matrix involved  is an $n\times n$ matrix. In the above example ${\cal E}_5$ is just 4! times the Hessian;
in the previous case,  ${\cal E}_4$ is just 6 times the sum of the four 3x3 Hessians formed by the second derivatives with respect to the four
possible subsets of three  of the variables $x_1,\ x_2, \ x_3,\ x_4$. Likewise ${\cal E}_3$ is just 
\[{\cal E}_3 =3(\sum_{\mu=1..4}\sum_{j=1..4}(\pd_{\mu\mu}\pi\pd_{\nu\nu}\pi -\pd_{\mu}^{\nu}\pi\pd_{\nu}^{\mu}\pi))
 \]
These terms are just proportional to the coefficients in the expansion of the characteristic polynomial of the matrix (\ref{hessn}).
These equations are solved by a similar technique to that above for ${\cal E}_5=0$.
These sequences of Lagrangians and equations of  motion are effectively the same as those of \cite{trodden1}.
These results are easily understood as examples of the main results of \cite{faigov} as is the fact that the equations of motion starting with a 
square root Lagrangian are of similar form to those starting from the Klein Gordon Lagrangian. (see \cite{dbf}).
\subsection{Multi-Galileons,} The theory has already been extended to the case of many fields
 ${\pi^i}$ \cite{deser}\cite{ padilla2} \cite{trodden2}. From 
the point of view  presented here, just as the single-field case was a Kaluza-Klein reduction of the Universal Field Equation, so are the 
multi-Galileon fields a reduction of the multi-field UFE, presented in \cite{gov3},  by setting the derivatives of the fields 
in the $x_5$th direction to unity. 
  The multi-field UFE  permits a first order formalism;
The set of equations of hydrodynamic type;
\be \frac{\pd u^i}{\pd x_0} + \sum u^j\frac{\pd u^i}{\pd x_j}=0\label{fluid}
\ee
are just the equations of conservation of momentum in an inviscid, incompressible fluid. Differentiation with respect to $x_j,\ (j=0...n)$
gives a set of second order differential equations, and the eliminants of those together with (\ref{fluid}) are the multi-field equations. 
The equations (\ref{fluid}) admit an easy implicit solution. The details may be found in \cite{fai2}. Indeed, multi-field UFE , with up to
$n$ fields, $k$, say, may be obtained from the equations (\ref{fluid}) by setting  $u^j,\ j=k+1\dots n$ as functions of $u^i,\ i\dots k$,
and forming the eliminants of these equations together with their derivatives with respect to the independent variables.
In particular the single field UFI comes from the equation
\be \frac{\pd u}{\pd x_0} + \frac{\pd u}{\pd x_1} +\sum_{j=2\dots n} u^j(u)\frac{\pd u}{\pd x_j}=0\label{fluid1}
\ee     
Differentiation with respect to $x_k,\ k=0\dots n$ yields the equations
\be \frac{\pd^2 u}{\pd x_0,\pd x_k} + \frac{\pd^2 u}{\pd x_1\pd x_k} +\sum_{j=2\dots n} u^j(u)\frac{\pd^2 u}{\pd x_j\pd x_k}
+ \frac{\pd u}{\pd x_k}\left(\sum_{j=1\dots n} \frac{\pd}{\pd u} u^j(u)\frac{\pd u}{\pd x_j}\right)=0\label{fluid2}
\ee 
Elimination of the coefficients $\displaystyle{1,\ u^j(u), \sum_{j=1\dots n} \frac{\pd}{\pd u} u^m(u)\frac{\pd u}{\pd x_m}}$ yields the
UFI
\[ \det\left|\begin{array}{cc}
        0&\frac{\pd u}{\pd x_i}\\
       \frac{\pd u}{\pd x_j}&\frac{\pd^2 u}{\pd x_i\pd x_j}
 \end{array}\right| =0.
\]
The Galileon can be obtained by a similar process of elimination, starting from (\ref{fluid2}). Making the assumption $u^n=1$, and also
imposing the side condition 
\[ \sum_{m=1\dots n-1} \frac{\pd u^m(u)}{\pd u}\frac{\pd u}{\pd x_m}+\frac{\pd u^n(u)}{\pd u}\,=\,0\]
the last equation of (\ref{fluid2}) vanishes and the eliminant is the  determinant of the $n\times n$ matrix $\frac{\pd^2 u}{\pd {i}\pd {j}} $  to zero.
Thus the Galileon also has roots in fluid dynamics. 
\section{Conclusions}
The Galileon equations can be viewed as  arising from  a Kaluza Klein reduction of the universal field equation,
(or more prosaically, as Chaundy has called it \cite{chaundy}, the bordered determinant) by setting the derivative
 of the fields with respect to the fifth co-ordinate to unity. This universal field equation admits a first order formalism in 
terms of an equation of hydrodynamic type, and is solved implicitly. The single field Galileon equation is 
a sum of Hessians set to zero and also admits implicit solutions. The universal field equation arises in the theory of
developable surfaces, and has also connections with Dirac-Born-Infeld theory \cite{bf}\cite{bf2}. Very recently a paper has 
appeared \cite{Acoleyen} showing that the Galileon actions and their covariant generalisations can also be obtained as a 
Kaluza Klein compactification of higher dimensional Lovelock Gravity. 

 \section{Acknowledgment} I should like to acknowledge Jan Govaerts for useful comments.

\end{document}